\documentclass{elsarticle}
\usepackage{graphicx}
\usepackage{hyperref}
\usepackage{color}
\usepackage{lineno}
\biboptions{numbers,sort&compress}
\usepackage[margin=3cm]{geometry}
\makeatletter
\def\ps@pprintTitle{%
 \let\@oddhead\@empty
 \let\@evenhead\@empty
 \def\@oddfoot{}%
 \let\@evenfoot\@oddfoot}
\makeatother

\begin{document}

\begin{frontmatter}

\title{A beam-beam monitoring detector for the MPD experiment at NICA}

\author[icn_unam]{Mauricio Alvarado}
\author[icn_unam]{Alejandro Ayala}
\author[cinvestav]{Marco Alberto Ayala-Torres}
\author[icn_unam,AE]{Wolfgang Bietenholz}
\author[uas]{Isabel Dominguez}
\author[cinvestav]{Marcos Fontaine}
\author[fcfm_buap]{P. Gonz\'alez-Zamora}
\author[cinvestav]{Luis Manuel Monta\~no}
\author[fcfm_buap]{E. Moreno-Barbosa}
\author[icn_unam]{Miguel Enrique Pati\~no Salazar}
\author[fcfm_buap]{L. A. P. Moreno}
\author[uas]{P. A. Nieto-Mar\'in}
\author[fcfm_buap]{V. Z. Reyna Ortiz}
\author[fcfm_buap]{M. Rodr\'iguez-Cahuantzi}
\author[fcfm_buap]{G. Tejeda-Mu\~noz}
\author[UCol]{Maria Elena Tejeda-Yeomans\corref{cor1}}
\cortext[cor1]{Principal corresponding author}
\ead{matejeda@ucol.mx}
\author[fcfm_buap]{A. Villatoro-Tello}
\author[cinvestav,fcfm_buap,conacyt]{C. H. Zepeda Fern\'andez}
\address[icn_unam]{Instituto de Ciencias Nucleares, Universidad Nacional Aut\'onoma de M\'exico, Apartado Postal 70-543, CdMx 04510, Mexico}
\address[cinvestav]{Centro de Investigaci\'on y Estudios Avanzados del IPN, Apartado Postal 14-740, CdMx 07000, Mexico}
\address[AE]{Albert Einstein Center for Fundamental Physics, Institute for Theoretical Physics, University of Bern, Sidlerstrasse 5, CH-3012 Bern, Switzerland}
\address[uas]{Facultad de Ciencias F\'isico-Matem\'aticas,  Universidad Aut\'onoma de Sinaloa, Av. de las
Am\'ericas y Blvd. Universitarios, Cd. Universitaria, CP 80000, Cln, Sinaloa, Mexico}
\address[fcfm_buap]{Facultad de Ciencias F\'isico Matem\'aticas, Benem\'erita Universidad Aut\'onoma de Puebla, Av. San Claudio y 18 Sur, Edif. EMA3-231, Ciudad Universitaria 72570, Puebla, Mexico}
\address[conacyt]{C\'atedra CONACyT, 03940 Ciudad de M\'exico, Mexico}
\address[UCol]{Facultad de Ciencias - CUICBAS, Universidad de Colima, Bernal D\'iaz del Castillo No. 340, Col. Villas San Sebasti\'an, 28045 Colima, Mexico.}

\begin{abstract}

The Multi-Purpose Detector (MPD) is to be installed at the Nuclotron Ion Collider fAcility (NICA) of the Joint Institute for Nuclear Research (JINR). Its main goal is to study the phase diagram of the strongly interacting matter produced in heavy-ion collisions. These studies, while providing insight into the physics of heavy-ion collisions, are relevant for improving our understanding of the evolution of the early Universe and the formation of neutron stars. In order to extend the MPD trigger capabilities, we propose to include a high granularity beam-beam monitoring detector (BE-BE) to provide a level-0 trigger signal with an expected time resolution of 30 ps. This new detector will improve the determination of the reaction plane by the MPD experiment, a key measurement for flow studies that provides physics insight into the early stages of the reaction. In this work, we use simulated Au+Au collisions at NICA energies to show the potential of such a detector to determine the event plane resolution, providing further redundancy to the detectors originally considered for this purpose namely, the Fast Forward Detector (FFD) and the Hadron Calorimeter (HCAL). We also show our results for the time resolution studies of two prototype cells carried out at the T10 beam line at the CERN PS complex.

\end{abstract}

\begin{keyword} \texttt{plastic scintillator,} 
\texttt{silicon photon detector},
\texttt{beam monitor}, \texttt{MPD,} \texttt{NICA} \end{keyword}

\end{frontmatter}

\section{Introduction}

The Multi-Purpose Detector (MPD)~\cite{MPD1Slava} is an experimental array to be installed at one of the interaction points of the Nuclotron Ion Collider fAcility (NICA). It will consist of central and forward detectors for charged hadrons, leptons and photons that will be produced in collisions with center-of-mass energies ($\sqrt{s_{NN}}$) ranging from 4 to 11 GeV for Au+Au, and up to 27 GeV for p+p. During the first stage of NICA operations, the MPD group will focus on studies of particle yields and spectra, event-by-event fluctuations, collective flow for identified hadrons and electromagnetic probes. During the second stage, the MPD group will address total particle multiplicities, asymmetries and the production of dileptons, charmonia, soft photons and hyper-nuclei. In order to guarantee optimal data taking at the different stages, it would be advantageous to have a system that is capable to provide a level-0 trigger signal and to help determine important observables such as the reaction plane~\cite{mario2}, the collision centrality~\cite{mario1} as well as to help discriminate true events from beam-gas interaction events. 

An example of the above mentioned system is the V0 ALICE-LHC detector~\cite{mario3} which was built using both BC404 and BC408 scintillating plastic. With this detector, the collected light is transported to photo-multipliers (PMTs) using optic fibers. At the LHC energies, this system has served very well during Run 1 and Run 2 as the main centrality detector for ALICE. For the upcoming LHC Run 3, with the expected six-fold increase in Pb-Pb luminosity and 50 kHz interaction rate, ALICE will use a new Fast Interaction Trigger (FIT)~\cite{mario4}. FIT will include a large ring of scintillating plastic with clear optical fibers and PMTs and two arrays of quartz radiators coupled to MCP sensors providing the time resolution for Pb-Pb collisions of 30 ps or less.

For the first NICA-MPD run scheduled for 2021 (with a limited luminosity of $5\times 10^{25}$ cm$^{-2}$ s$^{-1}$, and a rate in Au+Au collisions at the maximum energy up to 7 KHz~\cite{MPD-lumi}) it would be convenient to have a detector able to identify true collision events with a time resolution around 30 ps, to provide a wake up signal for the Time of Flight (TOF) in low multiplicity events, such as those coming from p+p or p+A reactions.

In this work, we describe the proposed geometry for a beam-beam monitoring detector (BE-BE) for the MPD, much in the spirit of the beam-beam counter (BBC) that was initially described in the \textit{Conceptual Design Report of the MPD}~\cite{cdrMPD}. We expect our proposed detector to become important for the trigger signal as well as for the centrality and event-plane resolution measurements in A+A collisions. Under certain colliding conditions, our detector would provide additional insight in p+A and p+p collisions, so it complements the Fast Forward Detector (FFD) and the Hadron Calorimeter (HCAL). Unlike the V0 and FIT, BE-BE's cells will be directly coupled to Silicon Photo-Multipliers (SiPM). The geometry we propose is closely related to the one used by the PHENIX-RHIC experiment~\cite{mario5}. We also discuss the potential for physics studies that can be carried out using the BE-BE in the analysis chain. Finally, we show results for time resolution\footnote{We mean the intrinsic resolution of the detector. Due to the limited
conditions of the beam test, we estimated the time resolution of the trigger and we used the FEE
time resolution reported by the ALICE Collaboration~\cite{mario3}. Currently, we are working on improving
our prototype design and further tests will allow us to estimate the final time resolution for the
complete BE-BE detector chain: cable length, signal treatment of the SiPM fast output, FEE
distortions, etc.} studies of two BE-BE cell prototypes tested at the T10 beam line at the CERN PS complex.

\section{BE-BE detector concept}

\begin{figure}
\centering\includegraphics[width=0.5\linewidth]{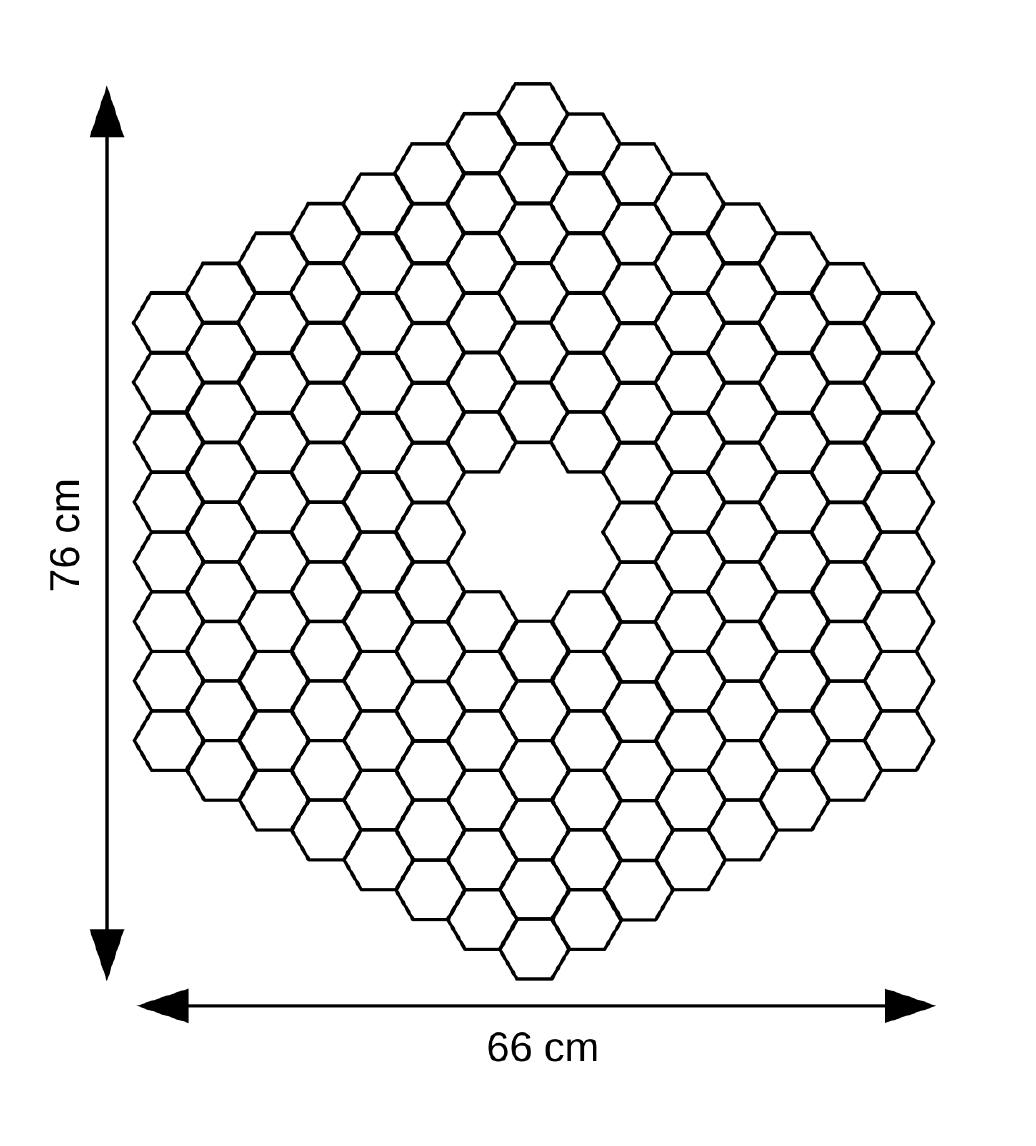}
\caption{One of the hodoscopes of the BE-BE design geometry as rendered by the MPD offline environment.  Each of the 162 hexagonal cells is made of plastic scintillator 5 cm high and 2 cm wide, which implies a size of $66~{\rm cm} \times 76~{\rm cm}$. The photons, which are emitted after charged particles go through the detector, are collected with light sensors.}
\label{fig:BBGeometry}
\end{figure}

The planned MPD beam-beam monitoring system must satisfy several requirements to be considered as a level-0 trigger. One of its crucial capabilities is the time resolution which, in order to serve as a trigger for TOF detector in low multiplicity collisions (p+p and p+A), is expected to be around 30 ps. To fulfill the trigger requirements, we propose that the BE-BE geometry consists of two hodoscope detectors, each located 2 meters away from the interaction point, at opposite sides. Each proposed detector consists of an array of 162 hexagonal plastic scintillator cells arranged in six concentric rings, as depicted in Fig.~\ref{fig:BBGeometry}. It covers a pseudo-rapidity range of $1.9< |\eta| < 3.97$. BE-BE will be useful to generate a trigger signal to identify and discriminate true from beam-gas events, either for minimum bias or with a given centrality, from background and beam-gas interactions. In addition, the BE-BE data can be used for the reconstruction of physical observables of interest in heavy-ion collisions such as the multiplicity of charged particles, collision centrality determination, event plane resolution and the absolute cross section determination for luminosity measurements. 

\section{Event plane resolution with the BE-BE}

The BE-BE aims to improve the MPD's determination of the reaction plane by providing  further  redundancy  to  the  detectors  originally  considered for this purpose namely, the Fast Forward Detector (FFD) and the Hadron Calorimeter (HCAL). The event-plane direction is a key measurement for flow studies which provides physics insight into the early stages of the reaction. This is useful to study the anisotropic flow of particles produced in heavy-ion collisions which is typically quantified by the coefficients in the Fourier decomposition of the azimuthal angular particle distribution~\cite{Voloshin:1994mz, Poskanzer:1998yz}. If the particle azimuthal angle is measured with respect to the direction of the reaction plane~\cite{PhysRevC.77.034904}, then the Fourier analysis leads to
\begin{eqnarray}\label{eq:equation1}
E \frac{dN}{d^3p} = \frac{1}{2\pi} \frac{dN}{p_{\mathrm{T}}dp_{\mathrm{T}}d\eta} \left\lbrace 1 + 2 \sum_{n=1}^{\infty} v_{n} (p_{\mathrm{T}},\eta) \cos\left[n(\varphi - \Psi_n) \right]\right\rbrace,
\end{eqnarray}
where $E$, $N$, $p$, $p_{\mathrm{T}}$, $\varphi$ and $\eta$ are the particle's energy, yield, total 3-momentum, transverse momentum, azimuthal angle and pseudo-rapidity, respectively. $\Psi_n$ is the reaction plane angle corresponding to the $n^{\mathrm{th}}$-order harmonic, $v_\mathrm{n}$. Experimentally, $\Psi_n$ can be determined using the sub-event correlation method discussed in Ref.~\cite{PhysRevC.77.034904}. 

Profiting from the high granularity of the BE-BE, we can resolve the event plane angle $\Psi^{BB}_n$ corresponding to the $n^{\mathrm{th}}$-order harmonic, using the reconstructed multiplicity provided by each hexagon cell of the hodoscope as follows~\cite{Voloshin:2008dg}
\begin{eqnarray} \label{eq:equation2}
\Psi^{BB}_n = \frac{1}{n} \, \tan^{-1} \left[ \sum_{i=1}^{m} w_i \, \sin (n\varphi_i){\mbox{\Huge{/}}} \sum_{i=1}^{m} w_i \, \cos (n\varphi_i)     \right], 
\label{EventPlane}
\end{eqnarray}
where $w_i$ is the multiplicity measured in the $i$-th cell, $m$ is the total number of BE-BE cells and $\varphi_i$ is the $i$th-cell's azimuthal angle measured from the center of the hodoscope to the cell centroid.

\begin{figure}[!h]
\centering\includegraphics[width=1\linewidth]{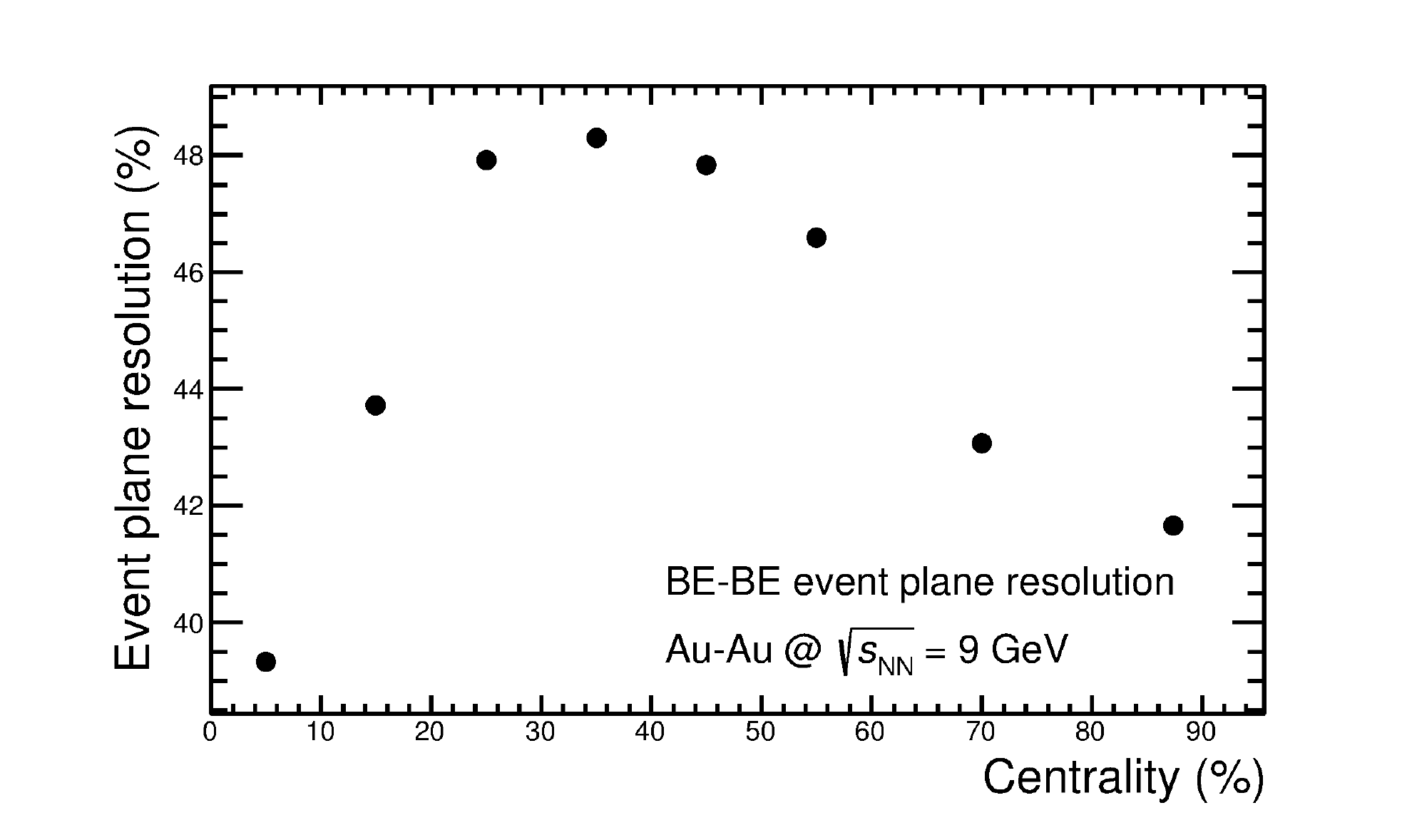}
\caption{Estimated event plane resolution using the BE-BE.}
\label{fig:BBEventPlane}
\end{figure}

To estimate the event plane resolution with the proposed BE-BE detector geometry, we simulated 95000 minimum bias Au+Au collision events at $\sqrt{s_{NN}}=$ 9 GeV in the centrality range of 0-90\%. The event generation was done with UrQMD~\cite{Bass:1998ca,Bleicher:1999xi}, which includes multiple particle interactions, the excitation and fragmentation of colour strings and the formation and decay of hadron resonances, in the simulation of p+p, p+A and A+A collisions. We used the MPD-ROOT offline framework~\cite{MPDROOT} to include the MPD-TPC~\cite{1748-0221-9-09-C09036} detector and the BE-BE. The produced particles were propagated through the detectors using GEANT-3 as transport package. The multiplicity per cell, $w_i$, was estimated at hit-level and the event plane resolution with the BE-BE detector for $n=1$ was computed as~\cite{Voloshin:2008dg}
\begin{equation}\label{eq:equation3}
\Big<\cos\Big(n\times (\Psi^{BB}_n-\Psi^{MC}_n)\Big)\Big >
\end{equation}
where $\Psi^{MC}_n$ is the true value given by the Monte Carlo for the $n$-th order harmonic. Figure~\ref{fig:BBEventPlane} shows the dependence of the event plane resolution with the centrality percentage for $n=1$. This effect has been also reported in Refs.~\cite{Abbas:2013taa, 1742-6596-742-1-012023,Ackermann:2000tr}. The BE-BE is capable to reach a maximum of the event plane resolution for a centrality range 25-45\% for Au+Au collisions at $\sqrt{s_{NN}}=9$ GeV. 

\section{Time resolution measurements for BE-BE prototypes}

\begin{figure}
\centering\includegraphics[width=0.7\linewidth]{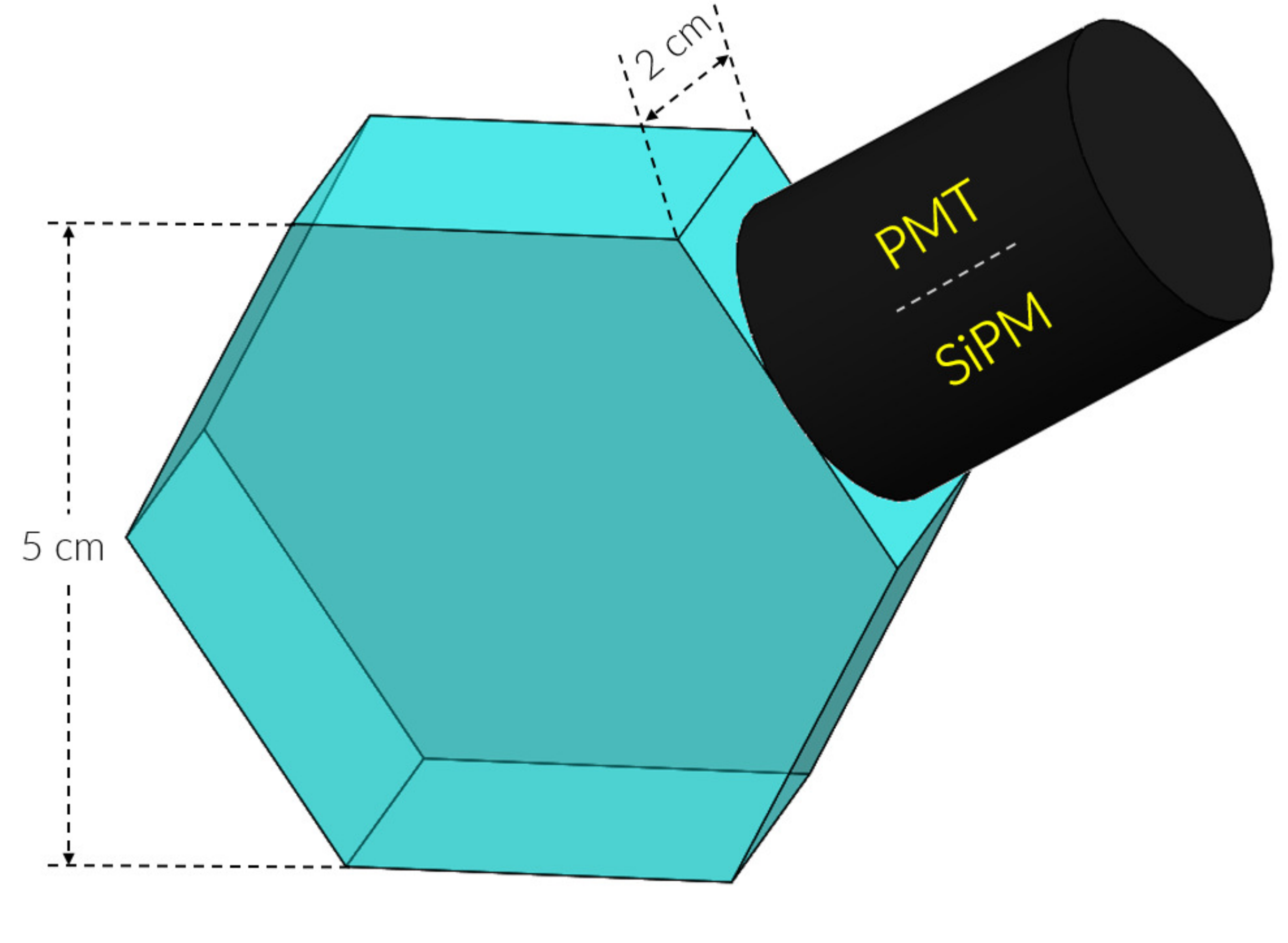}
\caption{Individual BE-BE cell prototype sketch. The BC-404 hexagonal cell is 5 cm high and 2 cm wide. Since the pion beam was aimed at the center of the cell, the light sensor was attached to one of the edges of the hexagonal cells to avoid possible damages caused by the beam. For the SiPM, the sensitive area is (6 mm)$^2$ whereas for the PMT it is the area of a 1 inch diameter circle.}
\label{fig:BBProto}
\end{figure}

In order to test the cell time resolution, we performed studies\footnote{The set-up described in Fig.~\ref{fig:BBProto} refers to the particular conditions during this first beam test. Under those conditions, we were protecting the PMT/SiPM devices from radiation burn-out. We are planning more tests in the near future with an array of configurations. Indeed, when we build the detectors, we will put the PMT/SiPM in one of the hexagonal faces of the cell, in order to create the arrangement shown schematically in  Fig.~\ref{fig:BBGeometry}} with two BE-BE cell prototypes made out of BC-404 plastic scintillators~\cite{bc404}. Each cell consists of a 5 cm high and 2 cm wide hexagon, as illustrated in Fig.~\ref{fig:BBProto}. To collect the light produced within the plastic scintillator, we used (1) a Hamamatsu PMT R6249~\cite{pmt} and (2) a SensL (C-60035-4P-EVB) SiPM~\cite{apd}. In both cases, the sensor was coupled to one of the edges of the hexagonal cell using a small amount of optical grease between the effective sensor light surface and the plastic scintillators, as depicted in Fig.~\ref{fig:BBProto}.

The machining of the hexagonal plastic scintillator cells was made with a high-pressure water jet cutter. The cells were then thoroughly polished. 
We used three layers of mylar and one layer of tyvek to wrap both BE-BE cell prototypes.

\begin{figure}
\centering\includegraphics[width=1\linewidth]{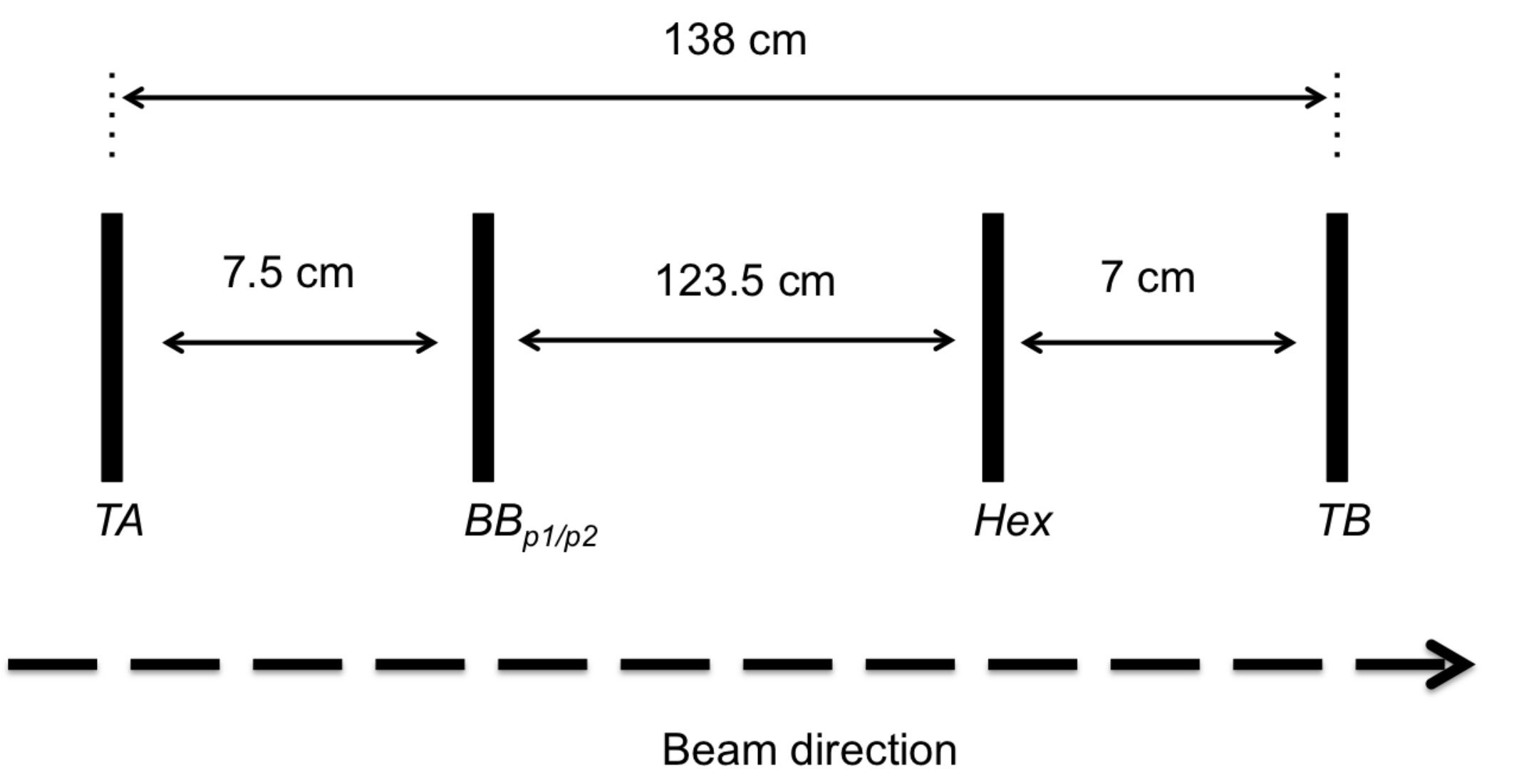}
\caption{Experimental setup in the pion beam of the T10 facility at CERN. The BE-BE hexagonal cell was located 7.5 cm from \emph{TA} and 123.5 cm from \emph{Hex}. The distance between \emph{Hex} and \emph{TB} was 7 cm. The pion beam direction is from left to right.}
\label{fig:BBTestArray}
\end{figure}

The pion beam energy was set to 5 GeV. For the trigger system, we used two thick scintillator paddles (\emph{TA}, \emph{TB}) and an additional hexagonal plastic scintillator detector (\emph{Hex}, used for acceptance) with the same dimensions of the BE-BE cell prototypes: $BB_{p1}$ (PMT R6249) and $BB_{p2}$ (SiPM). The evaluation of both prototypes was carried out in the pion beam test facility T10 at CERN~\cite{t10}. The complete experimental setup is shown schematically in Fig.~\ref{fig:BBTestArray}.

For the data acquisition system (DAQ), we used the front end electronics ($FEE$) developed for the ALICE V0 detector~\cite{mario3}, that reported 100 ps as its time resolution. This system provides useful trigger flags for the coincidence of two signals generated from individual plastic scintillator detectors (\emph{TA} and \emph{TB}). The time measurement is carried out with a high-performance time-to-digital converter chip (HPTDC)~\cite{HPTDC} with a global $FEE$ time resolution of 100 ps ($\sigma_{FEE}$ hereafter). Each of the signals from our experimental setup was connected to a high bandwidth pre-amplifier signal array (PASA). Each PASA has 2 outputs, one with a gain of $\times 1$ used for charge measurements and a second one with a gain of $\times 10$ used for triggering and time measurement~\cite{ZOCCARATO201190}.

The charge integration for all the connected signals is made every 25 ns. This is achieved with no dead time using two integrators working alternatively. The digital conversion is made with an analog-to-digital converter (ADC) of 20 MHz 10-Bit (AD9201). The signal which resets the integrators and the control of the digital conversion is driven through a field-programmable gate array (FPGA).

For the trigger generation and time measurement, an amplified version of the pulse is used. This pulse passes through a fast discriminator with a programmable discrimination level and it is measured with an HPTDC. With the help of a programmable delay based on the use of two shift registers, a window for time measurement can be set with a precision of 10 ps. For our test, the trigger condition was set as the coincidence (logical AND) between the two hodoscopes, \emph{TA} and \emph{TB}, within a time window of 25 ns.

\subsection{Data analysis}

\subsubsection{BE-BE cell prototype coupled to a Hamamatsu PMT R6249} 
\label{sec:BBPMT}

In order to estimate the BE-BE cell prototype time resolution ($\sigma_{BB_{p1}}$), we follow a similar procedure as has already been reported in the literature (see for example~\cite{timeres}). We plot and fit the time difference between measurements made by the \emph{TA} and $BB_{p1}$ detectors, along the efficiency curve of the PMT. The selection of valid events required a
coincidence signal between the \emph{TA}, \emph{TB}, \emph{Hex} and $BB_{p1}$ detectors.

\begin{figure}
\centering\includegraphics[width=1\linewidth]{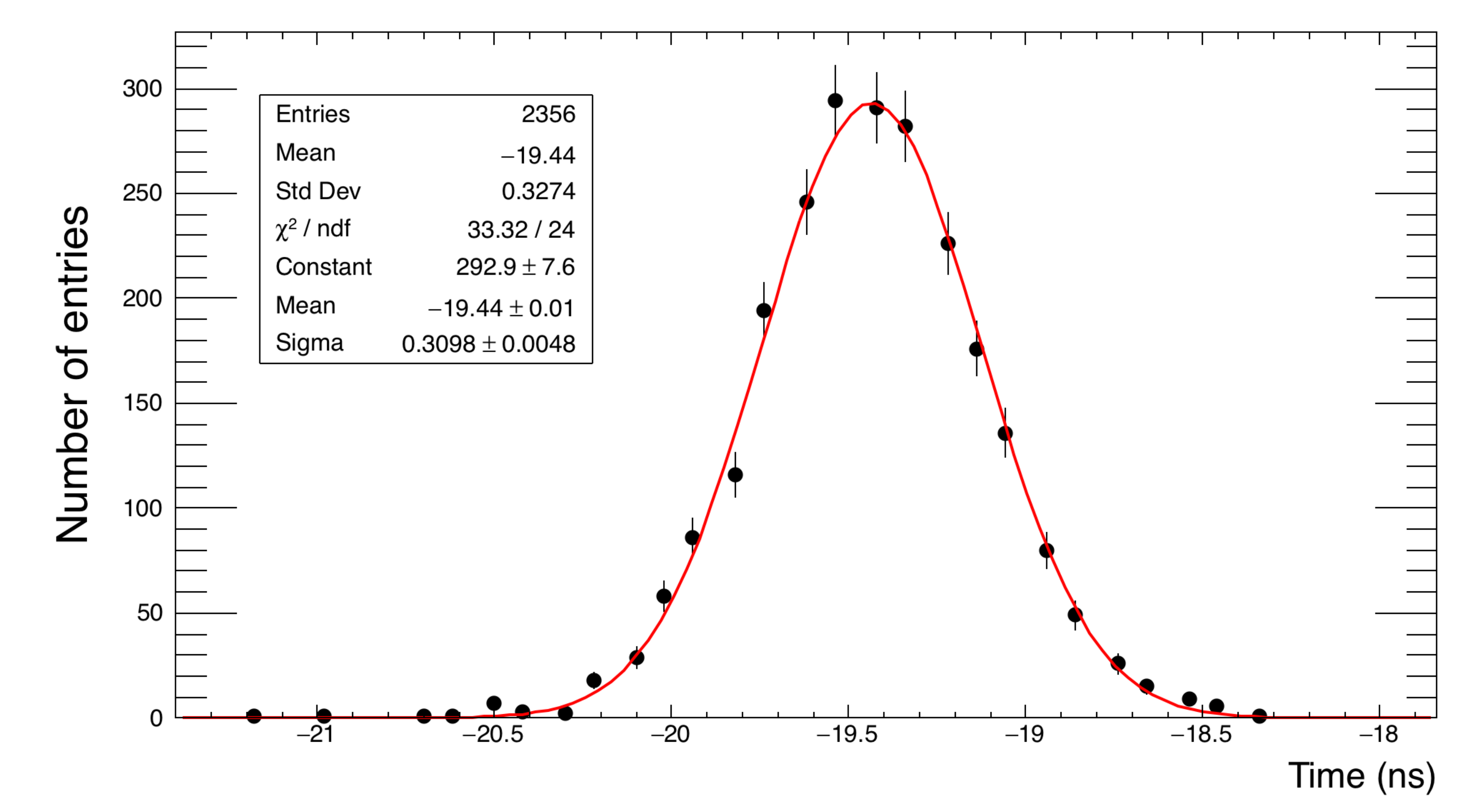}
\caption{Time difference between the measurement of \emph{TA} and $BB_{p1}$. The mean value around -19.44 ns comes from the time misalignment of each detector in the $FEE$. The PMT operational voltage was set to 700 V.}
\label{fig:DeltaTABBp1}
\end{figure}

\begin{figure}
\centering\includegraphics[width=1\linewidth]{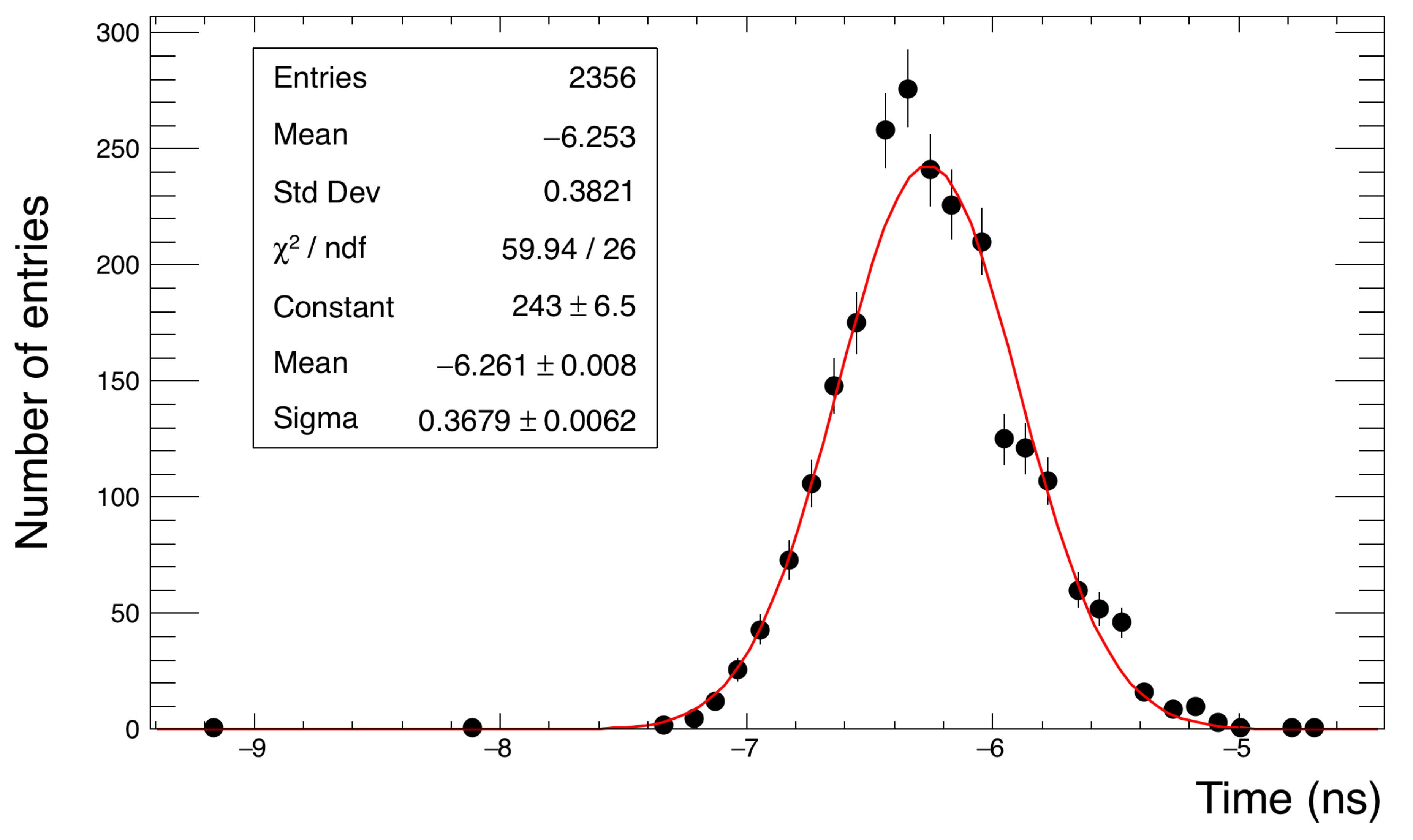}
\caption{Time difference between the measurements of \emph{TA} and \emph{TB}. The mean value around $-6.25$ ns comes from the misalignment in time of each detector in the $FEE$.}
\label{fig:DeltaTATB}
\end{figure}

As an example, Fig.~\ref{fig:DeltaTABBp1} shows the time difference between the measurements of \emph{TA} and $BB_{p1}$. The PMT's operational voltage was set to 700 V. In this case, we obtained the value $\sigma_{p1} = 0.309 \pm 0.005$  ns. Hereafter, the uncertainties are obtained from fits to data. We applied a standard Gaussian fit over all the range of experimental data given by the time difference between \emph{TA} and $BB_{p1}$ detectors. To analyze our experimental data we used the scientific software toolkit ROOT \cite{Brun:1997pa}.  
The $\sigma_{p1}$ value is related to the time resolution of the FEE and the \emph{TA} detector as
\begin{equation}
\label{eq:emc}
\sigma_{p1}^2 = \sigma_{BB_{p1}}^2 + \sigma_{TA}^2 + \sigma_{FEE}^2
\end{equation}
where $\sigma_{BB_{p1}}$, $\sigma_{TA}$ and $\sigma_{FEE}$ are the $BB_{p1}$, $TA$ and $FEE$ time resolutions, respectively. Furthermore, from the fit shown in Fig.~\ref{fig:DeltaTATB}, we obtained a value $\sigma_{T} =0.368 \pm 0.006$ ns, which is related to the detectors $TA$ and $TB$ as
$\sigma_T^2 = \sigma_{TA}^{2} + \sigma_{TB}^{2}$. Since \emph{TA} and \emph{TB} are identical detectors, we infer 
\begin{equation}
\label{eq:emc3}
\sigma_{TA} = \sigma_{TB} =\frac{\sigma_{T}}{\sqrt{2}}.  
\end{equation}

\begin{figure}[!h]\label{figure5}
\centering\includegraphics[width=1\linewidth]{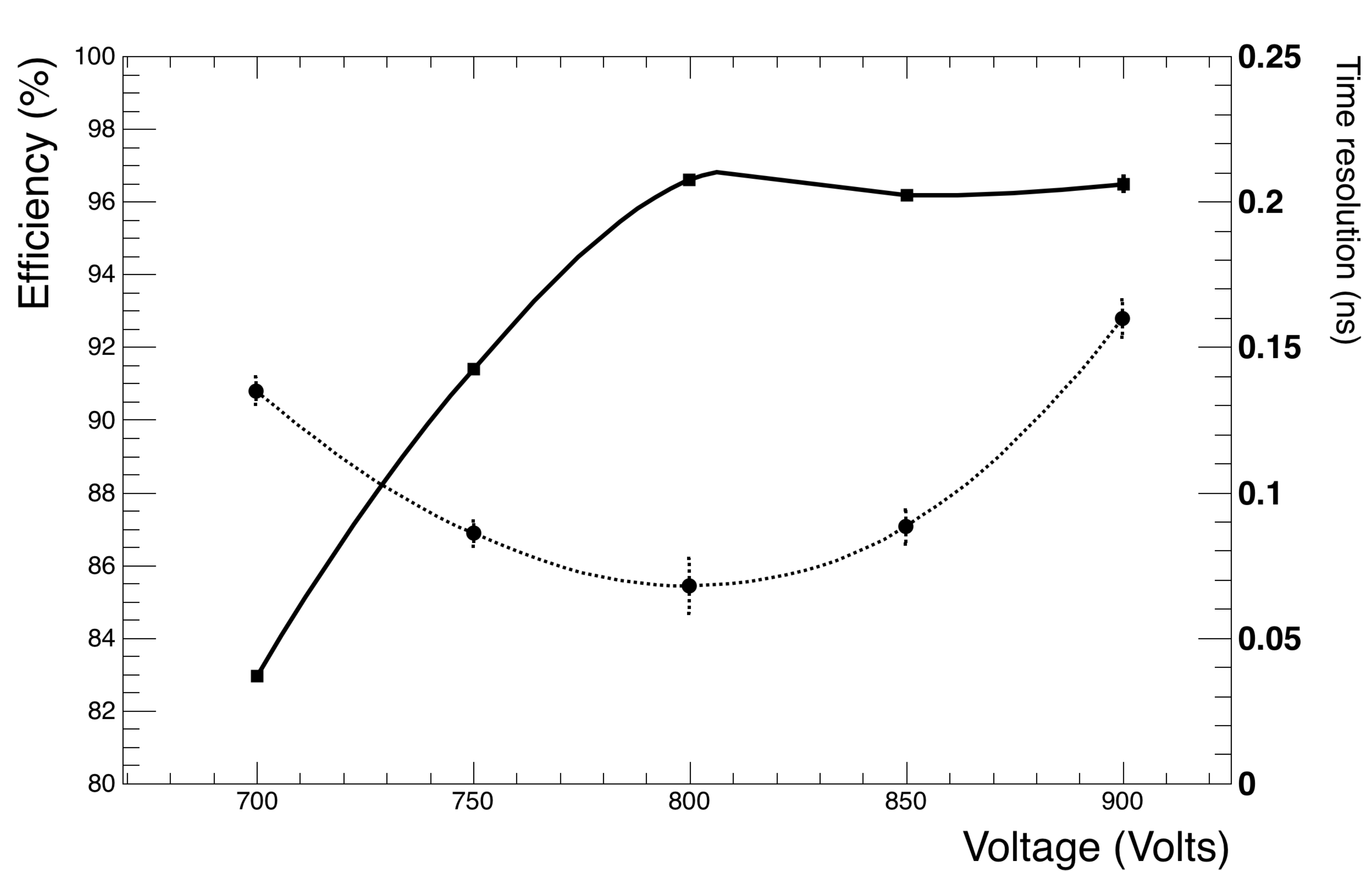}
\caption{Efficiency curve (continuous line) and time resolution (dotted line) of the $BB_{p1}$ detector as a function of the operational voltage for the Hamamatsu PMT R6249.}
\label{fig:BBTimeResolutionPMT}
\end{figure}

Following a similar procedure for the five values of the operating voltages for the PMT we used, we can extract the time resolution of the BE-BE cell prototype, $\sigma_{BB_{p1}}$, with respect to its efficiency curve shown in Fig.~\ref{fig:BBTimeResolutionPMT}. We observe that the cell detector prototype time resolution, $\sigma_{BB_{p1}}$, is better at the beginning of the plateau where the detector reached its maximum efficiency. The best time resolution value is $\sigma_{BB_{p1}} = 68 \pm 5$ ps for an operational voltage of 800 V. Notice that we achieve a bit less than 97\% efficiency. This is partly due to the 30 mV threshold operational voltage for the PMT coupled to the cell, which is rather high. Therefore some valid events for the \emph{TA} and \emph{TB} trigger coincidence are missed by the PMT-cell system. Table ~\ref{tab:TableBBTimeResolutionPMT} shows the different values for the BE-BE cell prototype time resolution for the different operational voltages used for the PMT.

\begin{table}
\centering
\resizebox{12cm}{!} {
\begin{tabular}{l l l l l l}
\hline
\textbf{Voltage (V)} & 700 & 750 & 800 & 850 & 900 \\
\hline
\textbf{Time resolution (ps)} & 135 $\pm$ 5 & 86 $\pm$ 4 & 68 $\pm$ 5 & 88 $\pm$ 6 & 160 $\pm$ 7 \\
\textbf{$\chi^2$/ndf} & 33.32/24 & 13.42/19 & 23.26/19 & 19.82/19 & 27.68/23 \\
\hline
\end{tabular}
}
\caption{Time resolution of the BE-BE cell prototype ($BB_{p1}$).}
\label{tab:TableBBTimeResolutionPMT}
\end{table}

\subsubsection{BE-BE cell prototype coupled to a SensL (C-60035-4P-EVB) Silicon Photomultiplier (SiPM)} 
\label{sec:BBAPD}

\begin{figure}
\centering\includegraphics[width=1\linewidth]{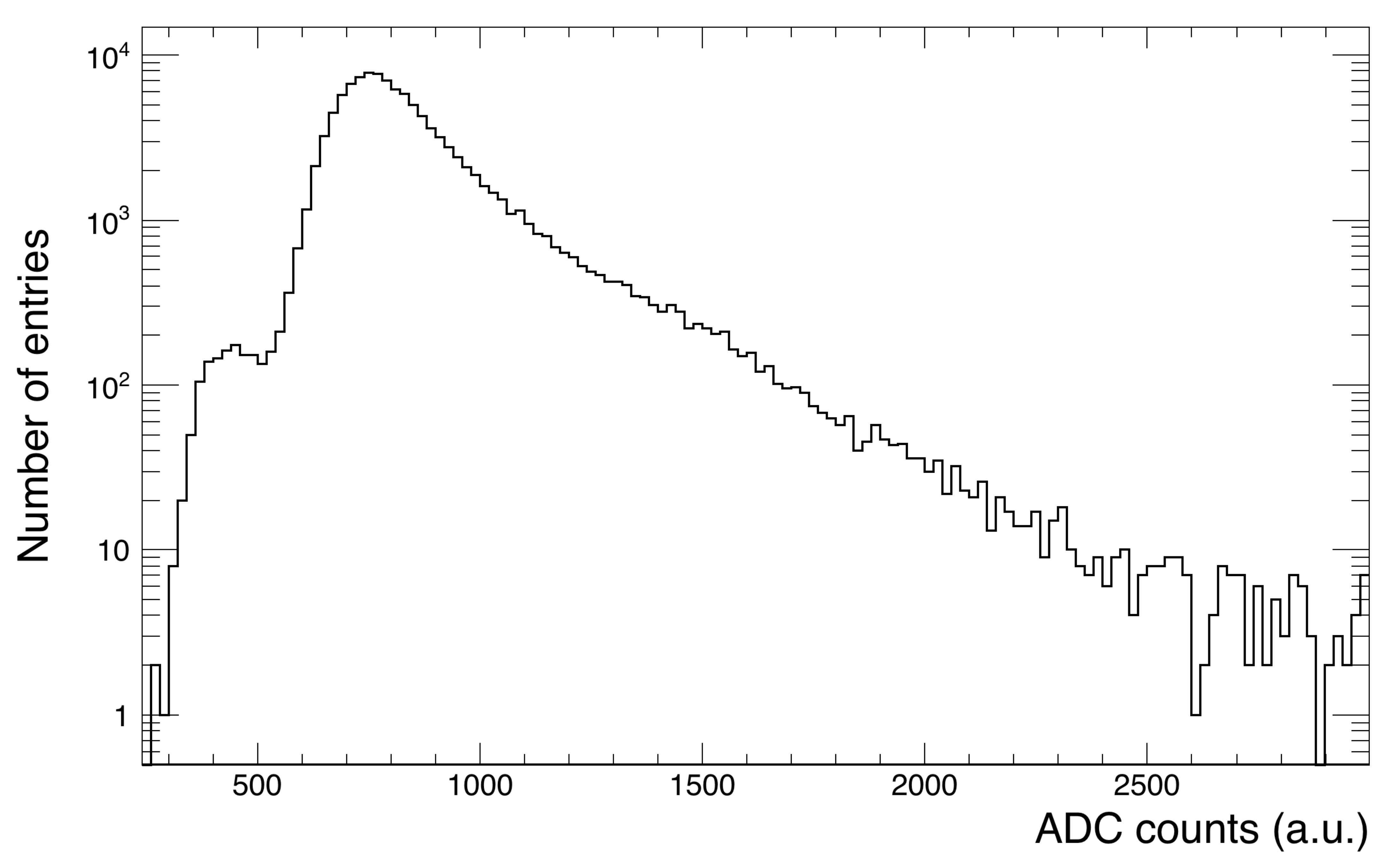}
\caption{ADC distribution of $BB_{p2}$ (SiPM-slow). The mean value is $860 \pm 7$ ADC counts in arbitrary units. The time resolution study was performed for different ADC ranges around the mean value of this distribution.}
\label{fig:BEBEP2ADCPlot}
\end{figure}

To estimate the time resolution for the BE-BE cell prototype detector coupled to a SiPM light sensor, $BB_{p2}$, we applied different selection criteria around the mean value of the charge distribution collected by the slow output of SiPM (SiPM-slow). The time measurement was performed with the signal provided by the SiPM's fast output (SiPM-fast). Figure~\ref{fig:BEBEP2ADCPlot} shows the ADC distribution of $BB_{p2}$ (SiPM-slow). The selection of valid events required a coincidence signal between \emph{TA}, \emph{TB}, \emph{Hex} and $BB_{p2}$ detectors. 
These selection criteria were required for both outputs of the SiPM light sensor, slow and fast, respectively.

\begin{figure}
\centering\includegraphics[width=1\linewidth]{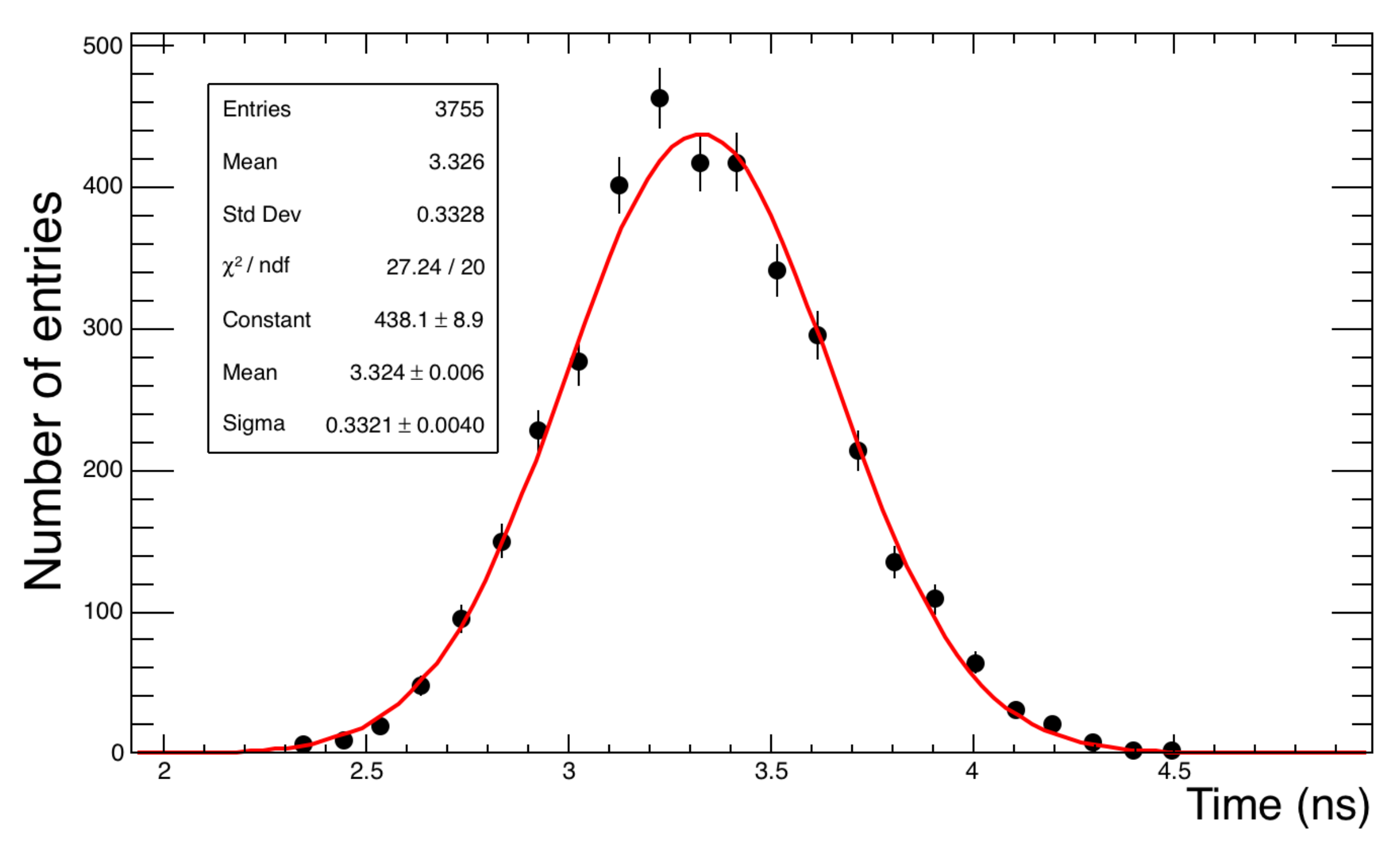}
\caption{Time difference between the measurement of \emph{TA} and $BB_{p2}$. The mean value around 3.32 ns comes from the misalignment in time of each detector in the $FEE$.}
\label{fig:TimeDifferenceTABBAPD}
\end{figure}

Figure~\ref{fig:TimeDifferenceTABBAPD} shows the time difference between $BB_{p2}$ and \emph{TA} for a range 850-870 in arbitrary units (a.u.) of the ADC. The fit results in a value of $\sigma_{p2} = 0.332 \pm 0.004$ ns. This value is related to the $FEE$, \emph{TA} and \emph{TB} time resolutions (recall $TA$ and $TB$ are identical detectors) by
\begin{equation}
\label{eq:emc5}
\sigma_{p2}^2 = \sigma_{BB_{p2}}^2 + \sigma_{TA}^2 + \sigma_{FEE}^2
\end{equation}
where $\sigma_{BB_{p2}}$ is the BE-BE cell prototype time resolution ($BB_{p2}$), $\sigma_{TA}$ is \emph{TA} reference trigger detector time resolution and $\sigma_{FEE}$ is the $FEE$ time resolution. The value of $\sigma_{TA}$ was obtained with the method described in Section ~\ref{sec:BBPMT}. In this case, we reach a time resolution of $\sigma_{BB_{p2}} = 83 \pm 4$ ps.

\begin{figure}
\centering\includegraphics[width=1\linewidth]{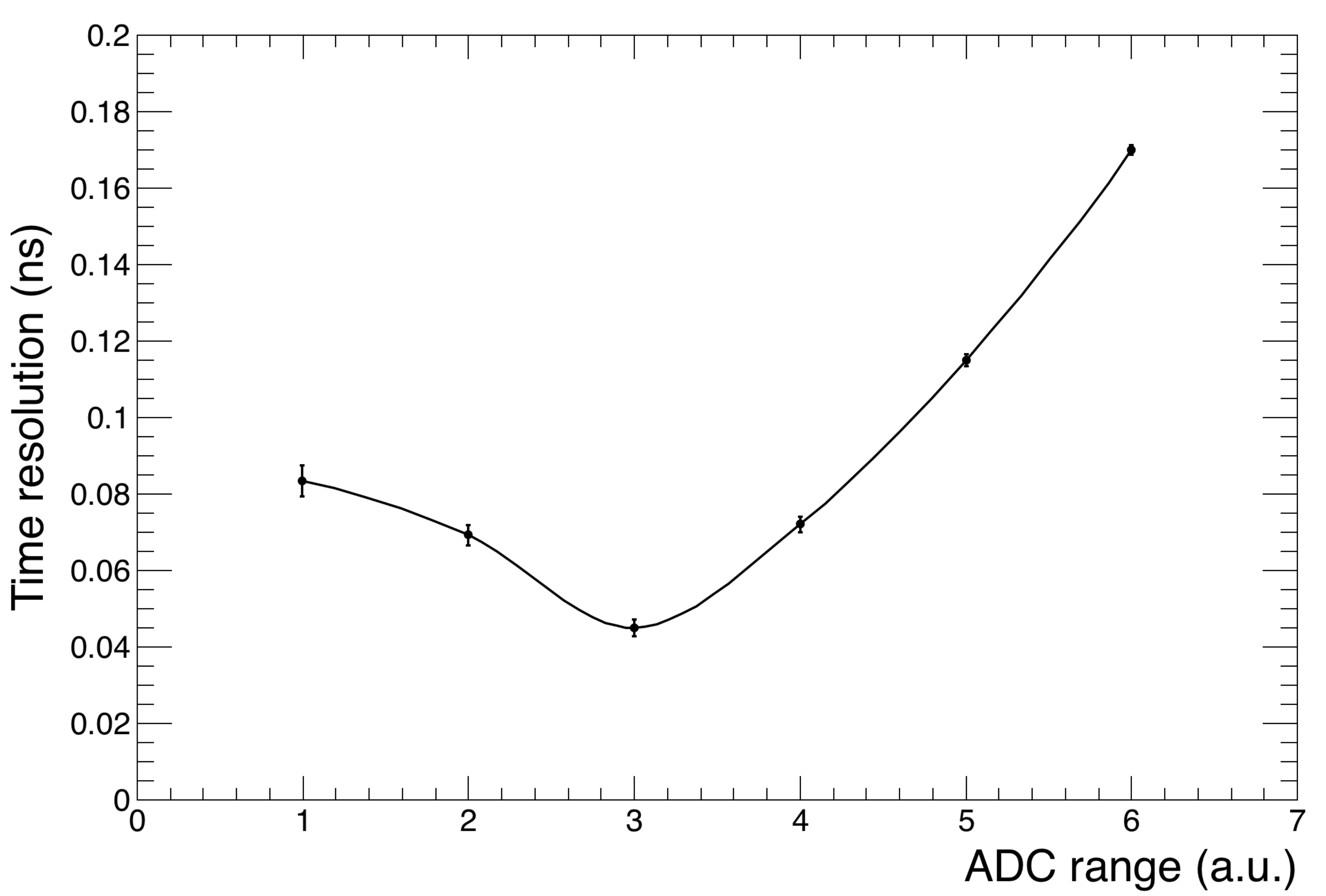}
\caption{$BB_{p2}$ prototype cell detector time resolution. The fast output of the SiPM was used to estimate the timing. The best time resolution is $45 \pm 2$ ps around the mean value of the ADC distribution of SiPM-slow signal. The horizontal axis refers to the labels which are assigned to the columns in Table~\ref{tab:TableTimeResolutionBBP2}.}
\label{fig:TimeResolutionBBP2}
\end{figure}

Figure~\ref{fig:TimeResolutionBBP2} shows the different ADC's ranges used to perform the $BB_{p2}$ cell prototype detector time resolution studies. These ranges are also listed in Table~\ref{tab:TableTimeResolutionBBP2}.

\begin{table}
\centering
\resizebox{12cm}{!} {
\begin{tabular}{l l l l l l l}
\hline
\textbf{ADC range (a.u.)} & 850-870 (1) & 840-880 (2) & 830-880 (3) & 830-890 (4) & 800-920 (5) & 700-900 (6) \\
\hline
\textbf{Time resolution (ps)} & 83 $\pm$ 4 & 69 $\pm$ 3 & 45 $\pm$ 2 & 72 $\pm$ 2 & 115 $\pm$ 2 & 170 $\pm$ 1 \\
\textbf{$\chi^2$/ndf} & 27.24/20 & 60.09/24 & 82.22/25 & 65.14/25 & 150.3/28 & 140.2/31 \\
\hline
\end{tabular}
}
\caption{Time resolution of the BE-BE cell prototype ($BB_{p2}$).}
\label{tab:TableTimeResolutionBBP2}
\end{table}

\subsection{Simulation of BE-BE cell prototype with Geant-4}
\begin{figure}
\centering\includegraphics[width=1\linewidth]{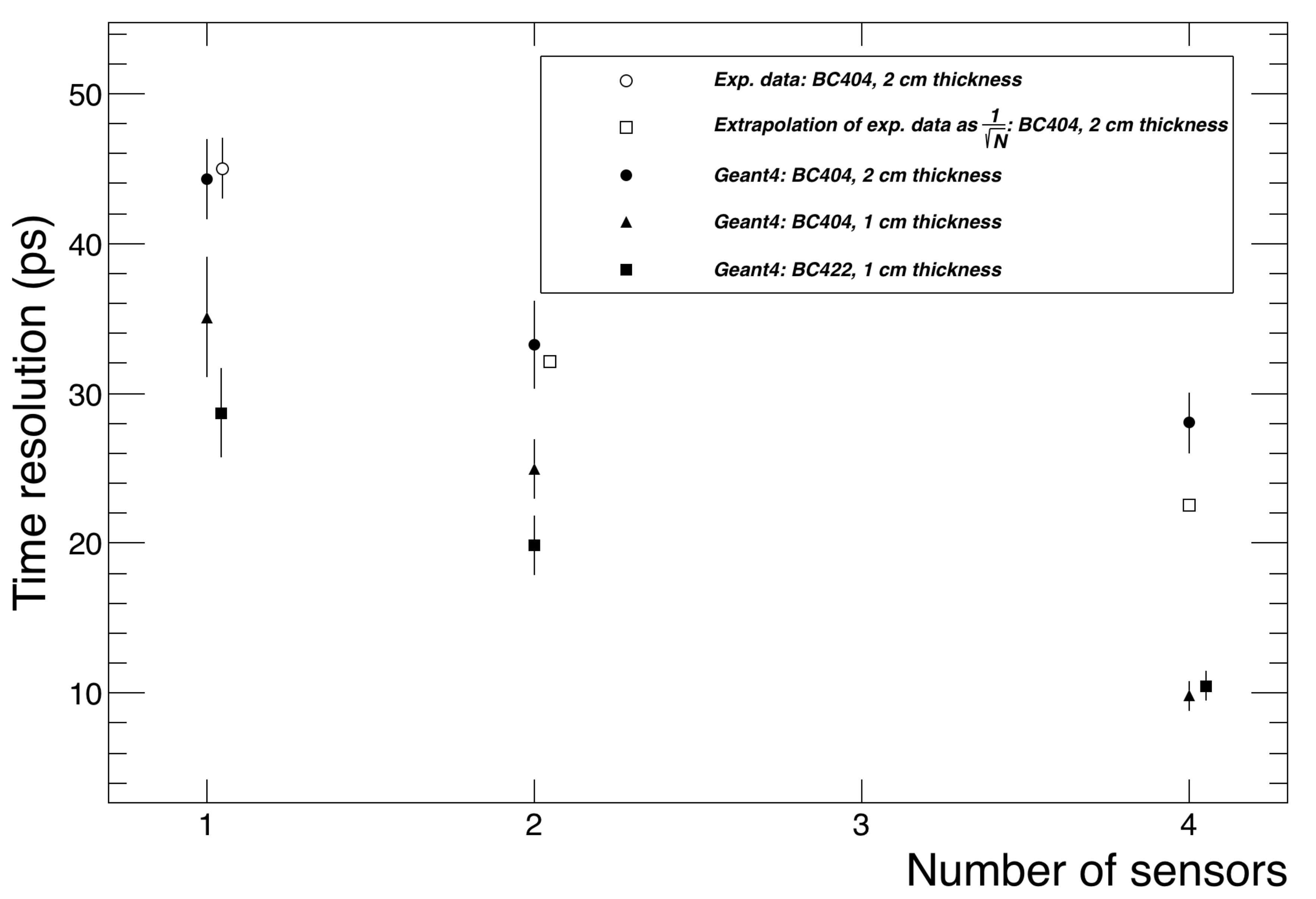}
\caption{$BB_{p2}$ prototype cell detector time resolution. The empty circle represents the experimental result ($\sigma_{BB_{p2}}$). The time resolution estimated with GEANT-4 is also shown: full circles (BC404, 2 cm thickness), full triangles (BC404, 1 cm thickness) and full squares (BC422, 1 cm thickness). For the number of sensors $N=2$ and $N=4$, the empty squares represent the extrapolated central value assuming a $1/\sqrt{N}$ behavior of the time resolution. For better visibility, some symbols are slightly shifted to the right.}
\label{fig:TimeResolutionBBGeant}
\end{figure}

To extend the time resolution studies with the $BB_{p2}$ detector, we performed simulations using the GEANT-4 toolkit simulation software \cite{AGOSTINELLI2003250}. We simulated the $BB_{p2}$ geometry taking as sensitive materials the BC404 and BC422 plastic scintillator, respectively. Two different thicknesses were considered for the hexagonal cell, 1 cm and 2 cm. For each of these sizes, we simulated 1, 2 and 4 scorers which represent the photosensors (SensL-SiPM). The scorers were coupled to the edges of the hexagonal cell. In each configuration, we simulated 200 pion events with an energy of 5 GeV striking the center of the hexagon. The time resolution from simulations was estimated from the fit of the time of flight of the emitted photons within the plastic scintillator that reach each scorer following a Landau distribution. Figure~\ref{fig:TimeResolutionBBGeant} shows a comparison of GEANT-4 simulations and experimental data from the beam test. We notice that for a hexagonal cell made of BC404, coupled with one light sensor, the simulations and the experimental data are in good agreement. Assuming that such a time resolution behaves as $1/\sqrt{N}$, where $N$ is the number of light sensors coupled to the plastic scintillator, the time resolution of a 5 cm high cell made of a BC404 hexagon coupled to two SiPM light sensors, will have a central value of 31 ps. GEANT-4 simulations suggest that such time resolution can be reached if we use a BC404 plastic scintillator 1 cm thick coupled to one light sensor or if we use a faster plastic scintillator such as BC422, coupled to one light sensor. The development of particle detectors with such a time resolution has been explored also in Refs.~\cite{Cattaneo:2014uya,Zhao:2016pja,HOISCHEN2011354}.

\begin{figure}
\centering\includegraphics[width=0.8\linewidth]{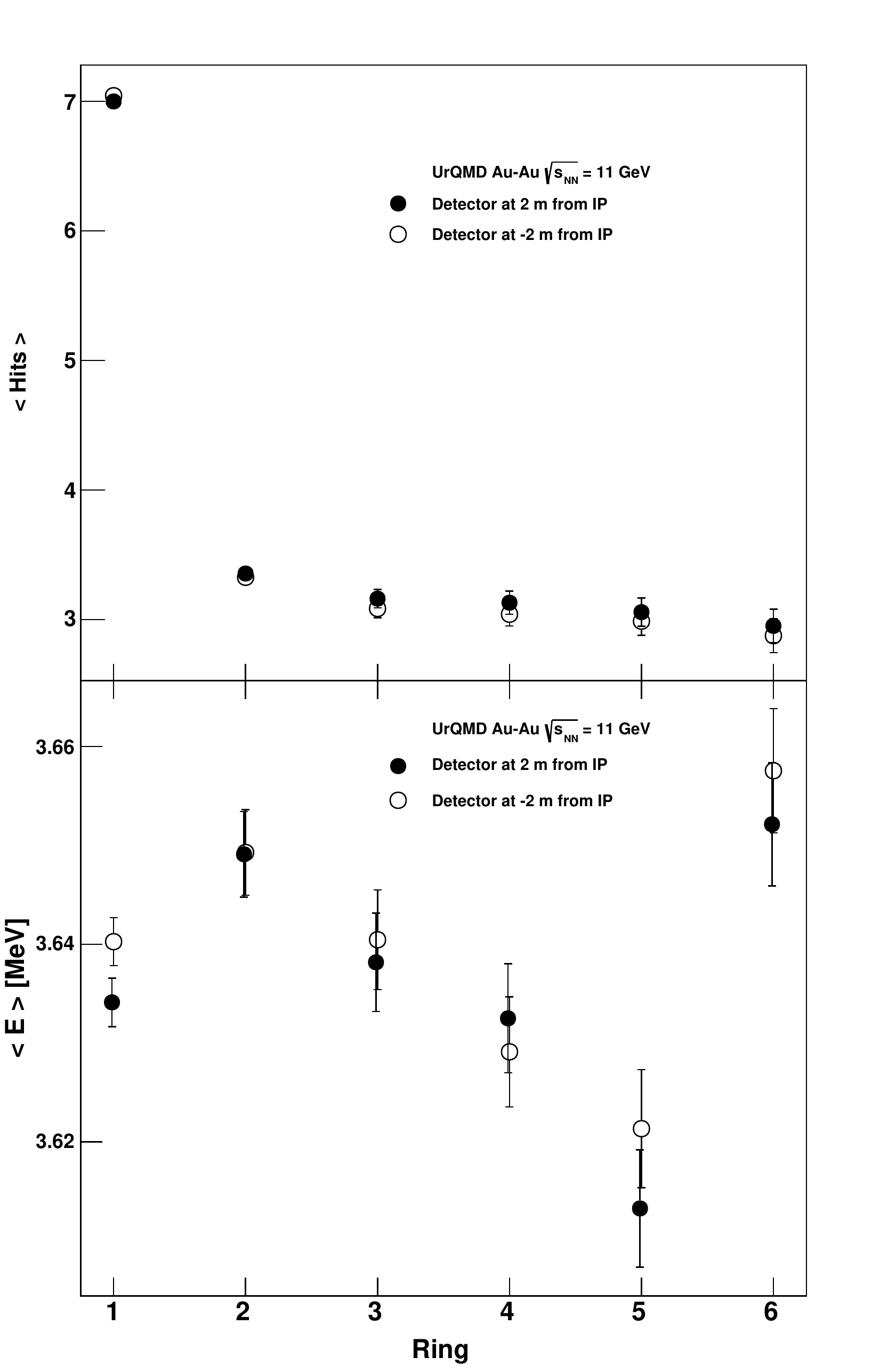}
\caption{Average number of hits per ring per event and average energy loss into the detector's active volume per ring per event for both sides of the detector, simulated using UrQMD and MPDROOT for minimum bias Au+Au collisions at $\sqrt{s_{NN}}$ = 11 GeV. We consider the 6 rings as per Fig.~\ref{fig:BBGeometry}, with 1 being the innermost ring and 6 being the outermost ring. The average number of hits per ring per event, that deposit an average energy well above the minimum ionizing energy, together with the efficient photon emission in this material, suggests an adequate trigger, that with an optimized geometry and distribution of light sensors could achieve the desired time resolution.}
\label{fig:mips}
\end{figure}

The 30 ps time resolution is necessary to determine a precise collision time for the TOF-based particle identification. This is particularly needed for low-multiplicity events. So, when implementing the future design modifications on our
detector we will improve the measurement of the final time
resolution by taking into account the complete BE-BE detector chain (cable length, signal treatment of the SiPM fast output, FEE distortions, etc.) and the number of charged particles penetrating the active volume of the detector with minimum ionizing energy (MIPs). Our simulations, performed for minimum bias Au+Au collisions at $\sqrt{s_{NN}}=11$ GeV, indicate (Fig.~\ref{fig:mips}) that, on average, 7 charged particles hit the active volume of the innermost ring of the detector. Therefore, the actual time resolution of the entire detector should be significantly better than a single-particle resolution measured for a single detector cell at the CERN T10 Proton Synchrotron facility.

This result, combined with the results for the photon emission produced by charged particles travelling through the scintillating plastic, shows that with an optimized geometry and distribution of light sensors, the system can serve as an adequate trigger.

\section{Conclusions}

In May 2018, we tested two cell detector prototypes that are intended to be used for the construction of the beam-beam monitoring detector for the MPD-NICA experiment at JINR. Both prototypes were made out of BC-404 plastic scintillator. For the light sensors we chose a Hamamatsu PMT R6249 and a SensL (C-60035-4P-EVB) SiPM. The test was carried out in the pion beam of the T10 facility at CERN. The DAQ, trigger and offline systems used during this beam test were provided by the forward trigger detector group of the ALICE-LHC experiment~\cite{ZOCCARATO201190}. These systems are a replica of those used during Run 1 (2009-2013) and Run 2 (2015-2018) data taking of the ALICE-LHC at CERN, therefore our measurements were made using this tested and well calibrated system. 

The best achieved time resolution was 68 $\pm$ 5 ps for the BC-404 hexagon coupled to the Hamamatsu PMT R6249 ($BB_{p1}$) and 45 $\pm$ 2 ps for the BC-404 hexagon coupled to the SensL ($BB_{p2}$) light sensor. Our results suggest that the desired time resolution for the beam monitor detector of the MPD/NICA experiment could be achieved with a proper optimization of the geometry and the number and distribution of SiPM attached to the cell. The achieved final time resolution depends also on the number of charged particles simultaneously penetrating the active volume of the detector. Our simulations show that we have a number of particles with average ionizing energy well above the minimum for these materials, that hit on average the active volume of the innermost ring of the detector. Therefore, we are confident that this system represents a
useful trigger, with appropriate geometry and well placed
light sensors.
Further tests using cosmic rays and
charged particles coming from radiation sources are being performed and will be reported elsewhere~\cite{MexNICA-RMF}. In both cases, the main issue in the MPD environment would be the design of an FEE capable of handling the signals coming from either the PMT or the SiPM light sensors to produce a level-0 trigger signal with a time resolution of 30 ps for the MPD-NICA. 

The results of the Au+Au simulations at $\sqrt{s_{NN}} = 9$ GeV show that with the proposed geometry for the beam-beam monitoring detector, it would be possible to reconstruct the reaction plane with a maximum resolution of 48\% for a centrality range between 25\% and 45\%. The BE-BE detector could provide valuable information to complement and to improve the reconstruction of the event plane in heavy-ion collisions at NICA energies with the MPD.

We are now in the process of studying each component of the
experimental setup that will ultimately be part of MPD-NICA. Ideally, we should estimate the
contribution of the FEE to the global time resolution obtained with the data from the beam test itself, as well
as from other sources, such as cable length, electronic noise, jitter, etc. Due
to the limited conditions of the beam test, we could not achieve a precise determination of the
time resolution of the FEE, so for the purpose of this analysis, we used the FEE time resolution
reported by the VZERO-ALICE group~\cite{mario3} cf.\ footnote 1. This assumption is susceptible to a number of variables, for instance the fact that we are using the pion beam of T10. So we estimated the uncertainty from the Gaussian fits to the experimental data. This is a ”best case
scenario” analysis and the ultimate goal of our collaboration is to perform new beam tests with
an improved prototype and analysis, where we would now have the available time to use the
facilities to collect data in such a way that we can obtain a global time resolution. This is work in progress and will be reported elsewhere.

\section*{Acknowledgments}

Support for this work has been received in part by Consejo Nacional de Ciencia y Tecnolog\'ia grant number 256494 and by UNAM-DGAPA-PAPIIT grant numbers IN107915 and AG100219. M.E.T-Y acknowledges the support provided by Universidad de Sonora in the initial stages of this work and the continuing support and commitment provided now by Universidad de Colima. M.R.C. thankfully acknowledges computer resources, technical advise and support provided by Laboratorio Nacional de Superc\'omputo del Sureste de
M\'exico (LNS), a member of the CONACYT national laboratories, with project No.53/Primavera 2017 and VIEP-BUAP grants 100524451-VIEP2018 and 100467555-VIEP2018. W.B.\ thanks for support by the European Research Council under the European Union's Seventh Framework Programme (FP7/2007-2013)/ERC grant agreement 339220.

The authors would like to thank Slava Golovatyuk and Eleazar Cuautle for the careful reading of this manuscript and for useful comments. We also acknowledge the use of the instrumented beam line and data acquisition system installed by the ACORDE/AD-ALICE groups for the tests performed at the T10 PS  beam at CERN. We acknowledge the technical support from Christoph Mayer and Mario Ivan Mart\'inez-Hern\'andez. We also thank ICAT-UNAM for allowing us the use of their facilities and for technical support.


\end{document}